\documentclass[aps,pre,twocolumn,superscriptaddress,longbibliography,nobibnotes,nourl,noeprint,nodoi]{revtex4-1}
\usepackage{amsmath,amssymb,wasysym,physics,siunitx}
\usepackage{xcolor,graphicx}
\usepackage[colorlinks=true,citecolor=blue,urlcolor=blue,linkcolor=black]{hyperref}

\newcommand{\Bo}[1]{B$^{(#1)}$}
\newcommand{\Ph}[1]{P$^{(#1)}$}
\newcommand{\rms}{\rm\scriptscriptstyle}

\AtBeginDocument{%
    \newwrite\bibnotes
    \def\bibnotesext{Notes.bib}
    \immediate\openout\bibnotes=\jobname\bibnotesext
    \immediate\write\bibnotes{@CONTROL{REVTEX41Control}}
    \immediate\write\bibnotes{@CONTROL{%
    apsrev41Control,author="08",editor="1",pages="1",title="0",year="1"}}
     \if@filesw
     \immediate\write\@auxout{\string\citation{apsrev41Control}}%
    \fi
}%

\begin{document} 

\title[]{Predicting conductivities of alkali borophosphate glasses based on site energy distributions derived from network former unit concentrations}

\author{Marco Bosi}
\affiliation{Universit\"{a}t Osnabr\"{u}ck, Fachbereich Physik, Barbarastra{\ss}e 7, D-49076 Osnabr\"uck, Germany}

\author{Philipp Maass}
\affiliation{Universit\"{a}t Osnabr\"{u}ck, Fachbereich Physik, Barbarastra{\ss}e 7, D-49076 Osnabr\"uck, Germany}

\date{November 6, 2021} 

\begin{abstract}
For ion transport in network glasses, it is a great
  challenge to predict conductivities specifically based on structural
  properties.  To this end it is necessary to gain an understanding of
  the energy landscape where the thermally activated hopping motion of
  the ions takes place. For alkali borophosphate glasses, a
  statistical mechanical approach was suggested to predict essential
  characteristics of the distribution of energies at the residence
  sites of the mobile alkali ions. The corresponding distribution of
  site energies was derived from the chemical units forming the glassy
  network. A hopping model based on the site energy landscape allowed
  to model the change of conductivity activation energies with the
  borate to phosphate mixing ratio. Here we refine and extend this
  general approach to cope with minimal local activation barriers and
  to calculate dc-conductivities without the need of performing
  extensive Monte-Carlo simulations. This calculation relies on the
  mapping of the many-body ion dynamics onto a network of local
  conductances derived from the elementary jump rates of the mobile
  ions. Application of the theoretical modelling to three series of
  alkali borophosphate glasses with the compositions
  $0.33$Li$_2$O$-0.67$[$x$B$_2$O$_3$-$(1\!-\!x)$P$_2$O$_5$],
  $0.35$Na$_2$O$-0.65$[$x$B$_2$O$_3$-$(1\!-\!x)$P$_2$O$_5$] and
  $0.4$Na$_2$O$-0.6$[$x$B$_2$O$_3$-$(1\!-\!x)$P$_2$O$_5$] shows good
  agreement with experimental data.
\end{abstract}

\maketitle

\section{Introduction}
\label{sec:introduction}

Solid electrolytes are used in chemical sensors \cite{Fergus:2011},
electrochromic devices \cite{Yoo/etal:2006-p}, optical wave guides
\cite{Tervonen/etal:2011}, supercapacitors
\cite{Samui/Sivaraman:2010}, and batteries \cite{Kim/etal:2015}.
Compared to liquid electrolytes, they have in general higher energy
and power densities, and their use in all-solid-state batteries allows
for a safer operation of electronic devices and electric cars.  A good
understanding of ion transport in solid electrolytes is needed to
further optimise the chemical and physical properties of these
materials for various applications.

A prerequisite for a material-specific understanding of the ion
transport is a knowledge of energy landscapes that describe how the
interaction energy of the mobile ions with the immobile constituents
of the host matrix varies in space.  In defective crystalline
materials, the energy landscapes may be determined by introducing
interaction parameters describing all relevant local configurations of
immobile atoms/ions in the environment of the mobile
ions. Corresponding interaction parameters can be determined by ab
initio methods \cite{Koettgen/etal:2018} or, if small in number, by
fitting to experimental results \cite{Meyer/etal:1996}.

In glassy electrolytes, it poses a particular challenge to gain
knowledge of the energy landscape because of the amorphous structure
of these materials. For understanding generic (non-material specific)
features of the ionic transport behaviour, it is justified to merely
assume that some distributions of site and/or barrier energies exist
\cite{Dyre/etal:2009}.  This has been done, for example, to understand
the origin of the mixed alkali effect \cite{Maass:1999-p, Hunt:1999-p}, 
of non-Arrhenius behaviour in fast ion conducting glasses \cite{Maass/etal:1996} 
and universal properties of conductivity spectra in ion conducting glasses 
\cite{Baranovskii/Cordes:1999, Dyre/Schroeder:2000, Porto/etal:2000, Dieterich/Maass:2002}.
For a material specific modelling, one could apply molecular dynamics (MD)
simulations \cite{Habasaki/Hiwatari:2004, Lammert/etal:2009, Vogel:2004} 
but it is not clear at present whether information on
the energy landscapes extracted from such simulations is
reliable. This is due to questions about the quality of force fields
and due to the problem that structures obtained from a cooling
protocol in simulations may not reflect the ones obtained in
experiments.

An attempt to derive energetically favourable diffusion paths of
mobile ions in oxide glasses based on experimental observations was
suggested by combining reverse Monte Carlo (RMC) modelling with a bond
valence analysis \cite{Adams/Swenson:2000, Adams/Swenson:2005}. In
that approach, amorphous structures are obtained by RMC modelling of
neutron and Xray diffraction data. Subsequently, preferred diffusion
paths for the mobile ions are identified by considering chemical
constraints (primarily minimal distances to network forming ions) and
an upper bound for deviations of the mobile ions' bond valence to
oxygen ions from an ideal value.  However, in an MD study of a lithium
silicate glass \cite{Mueller/etal:2007}, this approach failed to
identify correctly the ion sites in the glassy network, i.e.\ the
regions where mobile ions reside for a much longer time compared to
other regions.  Also, concentrations of network forming units (NFUs)
in mixed glass former glasses could not be correctly reproduced by the
RMC modelling of diffraction data \cite{Karlsson/etal:2015, Schuch/etal:2012}

A method for constructing site energy landscapes in ion conducting
network glasses was suggested in Ref.~\cite{Schuch/etal:2011}.  It is
based on the idea that the dominant contribution to the spatial
variation of the energy landscape originates from the counter charges
of the alkali ions. These counter charges are provided by charged NFUs
that are irregularly distributed in the glassy network. The method has
so far been applied to alkali borophosphate glasses, where in the
theoretical modelling an uncorrelated random spatial distribution of
the NFUs is assumed. Eight different types of NFU units were
identified by magic-spinning nuclear magnetic resonance (MAS-NMR
\cite{Silver/Bray:1958, Eckert:2018, Youngman:2018}) in these glasses
and the experimentally determined concentrations of NFUs could be well
explained by a statistical mechanical modelling
\cite{Schuch/etal:2011, Bosi/etal:2021}.

The site energy landscape constructed from the NFU concentration was
in the following used as an essential input for a hopping model of the
thermally activated ionic motion in the glassy network. Long-range
diffusion coefficients and their activation energies resulting from
this hopping model were first calculated by KMC simulations. The
simulated variation of the activation energies with the borate to
phosphate mixing ratio turned out to be in very good agreement with
experimental observations \cite{Schuch/etal:2011}.

In a recent study \cite{Bosi/etal:2021} it was shown that the activation energies can be
even derived without KMC simulations by applying the theory developed
in Ref.~\cite{Ambegaokar/etal:1971}.  We find it remarkable that one
can predict structural properties and a connection to ion transport
quantities in these complex disordered materials by rather simple
numerical calculations based on analytical theories, that means
without performing extensive simulations of the network structure and
ionic motion.

Here we refine and extend the analytical approach used in Ref.~\cite{Bosi/etal:2021}
and investigate whether not only activation energies can be successfully modelled but
also dc-con\-duc\-tiv\-i\-ties. To this end, we use the mapping of the
many-body hopping dynamics to a network of conductances introduced in
Ref.~\cite{Ambegaokar/etal:1971}. The conductivities are determined by
solving Kirchhoff's equations for the conductance network with the
Kron reduction method \cite{Doerfler/Bullo:2013}.  In addition we take
into account a minimal local activation energy for elementary ion
jumps. The theoretical modelling is applied to three series of alkali
borophosphate glasses and compared to experimental data.

\section{Site Energy Landscapes derived from modelling of MAS-NMR data}
\label{sec:site-energies}

\begin{figure}[t!]
\centering
\includegraphics[width=0.85\columnwidth]{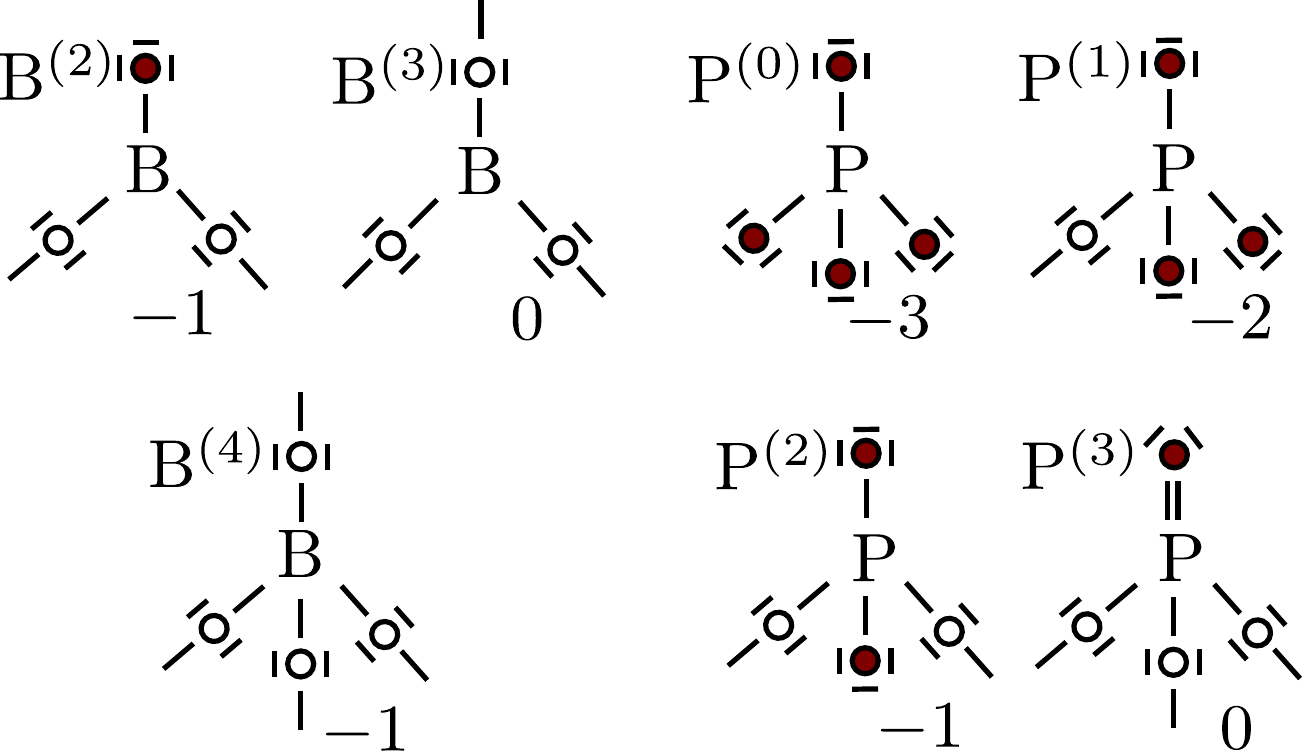}
\caption{Sketch of the units forming the glass network in alkali
  borophosphate glasses for molar alkali contents larger than 30\%
  (for lower alkali content, positively charged \Ph{4} units may
  appear \cite{Bosi/etal:2021}).  The numbers indicate the charge of
  the units and the nbOs of each unit are marked in red.}
\label{fig:nfu}
\end{figure}

The network of alkali borophosphate glasses is built by various borate
units \Bo{n} and phosphate units \Ph{m}, where $n$ and $m$ are
specifying the number of bridging oxygens (bOs). These units could be
identified by MAS-NMR \cite{Zielniok/etal:2007, Rinke/Eckert:2011, Larink/etal:2012}. 
As depicted in Fig.~\ref{fig:nfu},
there exists one negatively charged tetrahedral borate unit \Bo{4},
and two trigonal borate units \Bo{3} and \Bo{2}, where the \Bo{3} is
neutral and the \Bo{2} unit has one negative charge localised mainly
at its non-bridging oxygen (nbO, marked in red in
Fig.~\ref{fig:nfu}). By contrast, the negative charge of the \Bo{4}
can be considered as delocalised over all bOs. The phosphate units
\Ph{m} all have a tetragonal conformation and a negative charge
$(m-3)$ in units of the elementary charge. These
charges should be mainly localised at the nbOs and can be viewed to be
shared between these nbOs, i.e.\ between two nbOs for \Ph{2}, three
nbOs for \Ph{1}, and four nbOs for \Ph{0}. For borophosphate glasses
with low molar alkali content smaller than 30\%, positively charged
\Ph{4} units may occur \cite{Rinke/Eckert:2011}. In this study,
however, we are not considering such low alkali contents, which means
that the glass networks are formed by the units shown in
Fig.~\ref{fig:nfu}.

The concentrations of the respective NFU units vary with both the
borate to phosphate mixing ratio and the alkali content. A statistical
mechanical model was developed in our group to predict these
variations \cite{Schuch/etal:2011}.  Explicit analytical expressions
are given in Ref.~\cite{Bosi/etal:2021} if disregarding possible
disproportionation reactions between certain units.  The theoretical
predictions compare very well with the ones extracted from MAS-NMR
measurements. When generating site energy landscapes as described in
the following, we always used the NFU concentrations predicted by the
theory.

\begin{figure}[t!]
\centering
\includegraphics[width=0.95\columnwidth]{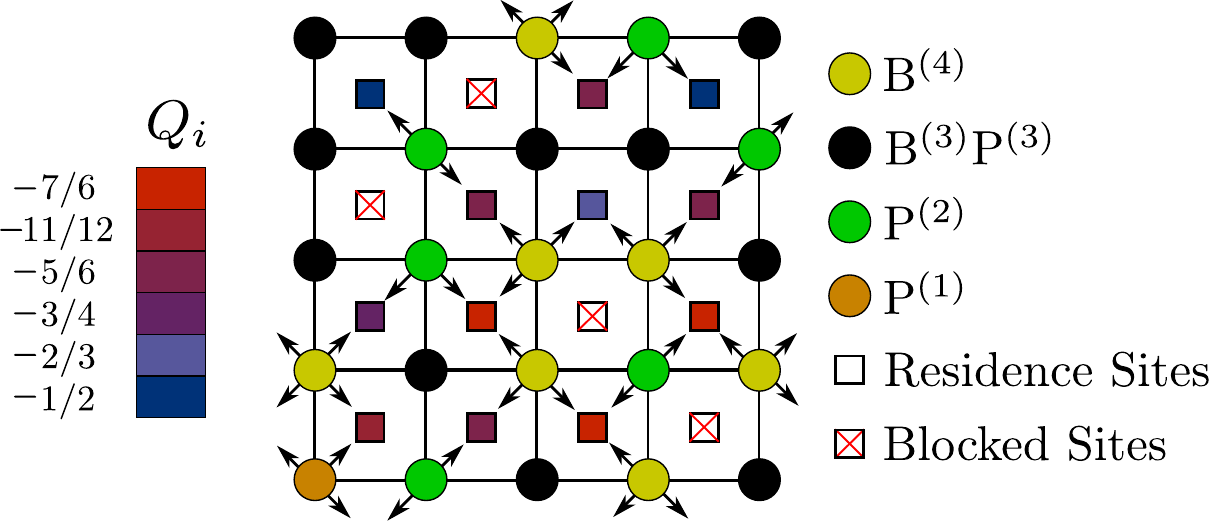}
\caption{Illustration of the generation of the site energy landscape.
  The NFUs are randomly distributed among the sites of a simple cubic
  lattice (here square lattice) and induce charges at the centres of
  the lattice cells (given in units of the elementary charge).  These
  centres are either residence sites of the mobile ions (``ion
  sites'') or blocked sites. At each ion site, a charge $Q_i$ is
  generated by the surrounding NFUs, which gives rise to a
  (non-smeared) energy $V_0Q_i$ at site $i$, see
  Eq.~\eqref{eq:epsi}. Specifically, an NFU with charge $q_\alpha$ and
  $k_\alpha$ nbOs adds a partial contribution $-q_\alpha/k_\alpha$ to
  $k_\alpha$ randomly selected neighbouring ion sites. A \Ph{2} unit,
  for example, gives a charge contribution -1/2 to two sites.  The
  delocalised charge -1 of a \Bo{4} unit induces an equal contribution
  to all its surrounding ion sites (indicated by circles). This
  contribution thus is $-1/z$, where $z$ is the number of non-blocked
  sites in the environment of the \Bo{4} unit (for the square lattice
  $z\le4$, while $z\le 8$ in the simple cubic lattice).  The partial
  charges induced by the NFUs are indicated by the arrows and they sum
  up to the total countercharge $Q_i$.  Different levels of the $Q_i$
  are represented by the color coding.}
\label{fig:illustration-e-generation}
\end{figure}

Knowing the NFU concentrations, site energy landscapes are created by
the procedure sketched in Fig.~\ref{fig:illustration-e-generation}
\cite{Schuch/etal:2011,Bosi/etal:2021}. For simplicity, we consider
one of the two interpenetrating simple cubic sublattices of a bcc
lattice as the positions of the NFUs and refer to it as the NFU
lattice. Each site of this NFU lattice is occupied by an NFU with a
probability given by its concentration weight without taking into
account spatial correlations.

The $N_{\rm tot}$ sites of the other simple cubic sublattice contain
the $N_{\rm acc}$ ion sites (residence sites of the mobile
ions).\footnote{This number $N_{\rm tot}$ is equal to the number of
  NFU sites in the construction here with the two simple cubic
  sublattices of the bcc lattice.  This construction requires that the
  number of alkali ions is not larger than that of the network forming
  cations, or, differently speaking, the molar alkali content must not
  exceed that of the network former cations. If glass compositions
  with higher alkali content shall be modelled, one can use other
  constructions of interpenetrating lattices.}  A fraction $f_0$ of
these accessible sites is empty. In MD simulations, $f_0$ was found to
be in the range of 5-10\%.  That $f_0$ must be small follows also from
theoretical considerations to explain internal friction measurements
\cite{Peibst/etal:2005, Maass/Peibst:2006-p}.  The $N_{\rm acc}$ sites
consist of $N$ occupied ion sites ($N$: number of ions) and $N_{\rm
  vac}$ vacant ion sites, $N_{\rm acc}=N+N_{\rm vac}$.  The remaining
sites in the respective sublattice are regarded as blocked
(non-accessible) for the mobile ions. Their number is $N_{\rm
  bl}=N_{\rm tot}-N_{\rm acc}$.  The blocking arises quite natural
here on stoichiometric reasons and leads to geometrically restricted
diffusion paths for the mobile ions, with unaccessible network regions
expanding when the alkali content becomes smaller.

For an alkali borophosphate glass with general composition
$yM_2$O-$(1\!-\!y)$[$x$B$_2$O$_3$-$(1\!-\!x)$P$_2$O$_5$] ($M$: alkali
ion, $y$: molar fraction of alkali oxide), the fraction of alkali ions
per network forming cation B and P is $y/(1-y)$. We thus have
$N/N_{\rm tot}=y/(1-y)$, because the total number $N_{\rm tot}$ of
sites in the two sublattices is equal in the construction considered
here.  Taking the value $f_0=0.05$ for the fraction $N_{\rm
  vac}/N_{\rm acc}$ of vacant ion sites accessible, the fraction
$f_{\rm bl}$ of blocked sites in the sublattice containing the ion
sites is
\begin{align}
f_{\rm bl}&=\frac{N_{\rm bl}}{N_{\rm tot}}=1-\frac{N_{\rm acc}}{N_{\rm tot}}
=1-\frac{N}{N_{\rm tot}}\frac{N_{\rm acc}}{N}\nonumber\\
&=1-\frac{y}{1-y}\frac{N_{\rm acc}}{N_{\rm acc}-N_{\rm vac}}=1-\frac{y}{(1-y)(1-f_0)}\,.
\end{align}
In the construction of the energy landscape, we first block a fraction $f_{\rm bl}$ of randomly chosen 
sites in the respective sublattice (corresponding to infinite site energies for these sites). 
Thereafter a countercharge $Q_i$ is assigned to each ion site $i$ 
as described in the caption of Fig.~\ref{fig:illustration-e-generation}.

The site energy at site $i$ is
\begin{equation}
\varepsilon_i=V_0(Q_i+\eta_i)\,,
\label{eq:epsi}
\end{equation}
where $V_0$ is an energy scale.
The $\eta_i$ are uncorrelated Gaussian random numbers with zero mean and variance 
$\sigma_\varepsilon^2$. They take into consideration that the network geometry is disordered 
(not a lattice) and that there will be fluctuations due to the Coulomb interaction between the mobile ions.
In the kinetics of the ions jumps discussed in the following Sec.~\ref{sec:hopping-model}, we use 
a simple model for the saddle point energies separating neighbouring energy minima.
Accordingly, the fluctuations of the $\eta_i$ are considered to take into account also fluctuations 
of these saddle point energies.

\section{Hopping Model for Ion Transport}
\label{sec:hopping-model}
As discussed above, we consider the Coulomb interaction between the
mobile ions to contribute an (average) amount to the site energies
$\varepsilon_i$, which is taken into account by the noise term in
Eq.~\eqref{eq:epsi}.  This corresponds to a treatment of this
long-range interaction by a mean-field approximation.\footnote{More
  precisely, we view it to give a mean-field contribution in the
  equivalent vacancy picture of the dynamics, as described in
  Ref.~\cite{Bosi/etal:2021}.}  The only interaction between the ions
is then a mutual site exclusion, which means that an ion site can be
occupied by at most one ion.

For a mobile ion to move from a site $i$ with site energy $\varepsilon_i$
to a vacant neighbouring sites $j$ with site energy $\varepsilon_j$, it
has to surmount a saddle point energy $\varepsilon_{ij}^{\rm s}$.  If the
differences between the saddle point and site energies are much larger
than the thermal energy $k_{\rm B}T$, these processes become jump-like
on a coarse-grained time scale. According to chemical kinetics,
the rate for a jump from site $i$ to a site $j$ is
\begin{equation}
w_{i\to j}^0=\nu\exp[-\beta(\varepsilon^{\rm s}_{ij}-\varepsilon_i)]\,,
\label{eq:w0-gen}
\end{equation}
where $\beta=1/(k_{\rm B}T)$ and $\nu$ is an attempt frequency. Values for $\nu$
are typically of the order of $10^{12}-10^{14}\,\si{Hz}$ corresponding to optical phonon frequencies.

The jump rates satisfy the detailed balance condition
$\exp(-\beta\varepsilon_i)w_{i\to j}^0=\exp(-\beta\varepsilon_j)w_{j\to
  i}^0$.  This holds under the condition that the initial site before
the jump is occupied and the target site of the jump is vacant.
Introducing occupation numbers $n_i$ for the ion sites $i$, where
$n_i=1$ if the site $i$ is occupied by a mobile ion and $n_i=0$
otherwise, we can write more generally $n_i(1-n_j)w_{i\to j}^{(0)}$
for the rate of a jump from site $i$ to $j$. It holds $\langle
n_i(1-n_j)w_{i\to j}^{(0)}\rangle_{\rm eq}=\langle n_i\rangle_{\rm
  eq}(1-\langle n_j\rangle_{\rm eq})w_{i\to j}^{(0)}$, where
$\langle\ldots\rangle$ denotes an equilibrium average in the grand
canonical ensemble. The mean occupation numbers are given by the Fermi
function
\begin{equation}
\langle n_i\rangle_{\rm eq}=\frac{1}{\exp(\beta(\varepsilon_i-\mu))+1}
\label{eq:nieq}
\end{equation}
with the chemical potential $\mu$ fixing the (mean) number of ions. We
note that the detailed balance condition implies $\langle
n_i\rangle_{\rm eq}(1-\langle n_j\rangle_{\rm eq})w_{i\to j}^{(0)}
=\langle n_j\rangle_{\rm eq}(1-\langle n_i\rangle_{\rm eq})w_{j\to
  i}^{(0)}$, which means that the stationary state of the hopping
dynamics given by the rates in Eq.~\eqref{eq:w0-gen} is the
equilibrium one characterised by the mean occupation numbers $\langle
n_i\rangle_{\rm eq}$.

The saddle point energies $\varepsilon_{ij}^{\rm s}$ must be larger
than both site energies $\varepsilon_i$ and $\varepsilon_j$,
$\varepsilon_{ij}^{\rm s}>\max(\varepsilon_i,\varepsilon_j)$. We thus
can write $\varepsilon_{ij}^{\rm
  s}=\max(\varepsilon_i,\varepsilon_j)+u_{ij}$, where $u_{ij}>0$ is
the lower barrier for the forward and backward jumps between sites $i$
to $j$.  These barriers will be distributed according to some
probability density $\psi_u(u)$ and may be correlated with the site
energies. For simplicity and due to missing information, we will use
the simplest ansatz here, where all $u_{ij}$ are equal, corresponding
to a delta-function $\psi(u)=\delta(u-u_0)$.  The jump rates in
Eq.~\eqref{eq:w0-gen} then become
\begin{equation}
w_{i\to j}^0=\nu e^{-\beta u_0}
\min(1,e^{-\beta(\varepsilon_j-\varepsilon_i)})\,.
\label{eq:metropolis-rates}
\end{equation}
This corresponds to a Metropolis form \cite{Metropolis/etal:1953} with
an effective, temperature dependent attempt frequency $\nu e^{-\beta
  u_0}$. The form has been used in earlier work with $u_0=0$
\cite{Schuch/etal:2011, Bosi/etal:2021}, but a barrier $u_0>0$ is
necessary for ensuring thermally activated local jump dynamics for all
possible transitions between neighbouring sites.

When introducing $u_0$ in the modelling, we are faced with one further
parameter that needs to be specified. To keep the number of parameters
as small as possible, we use a fixed value $u_0=0.1\,V_0$ here. This
value is motivated by the idea that the smallest barriers for jumps at
the glass transition temperature $T_{\rm g}$ should be of the order of
$k_{\rm B}T_{\rm g}$.  In Sec.~\ref{sec:conductivity-comparison} we
compare experimental results for the activation energy and
conductivity with theoretical predictions for the glasses with
compositions 0.33Li$_2$O$-0.67$[$x$B$_2$O$_3$-$(1\!-\!x)$P$_2$O$_5$],
0.35Na$_2$O$-0.65$[$x$B$_2$O$_3$-$(1\!-\!x)$P$_2$O$_5$], and
0.4Na$_2$O$-0.6$[$x$B$_2$O$_3$-$(1\!-\!x)$P$_2$O$_5$]. To fit the
experimentally determined activation energies, we find values for the
energy scale $V_0$ between $0.65\,\si{eV}$ and $0.93\,\si{eV}$.  The
glass transition temperatures of the various compositions are in the
range 500-775\,K. Taking a typical value of $T_{\rm
  g}\simeq700\,\si{K}$ corresponding to a $u_0\simeq k_{\rm B}T_{\rm
  g}\simeq0.06\,\si{eV}$, this yields $u_0/V_0$ values of about 1/10.

\section{Theory for Activation Energies and Conductivities}
\label{sec:conductivity-theory}

In the presence of an external electric field $\vb E$ acting on the
ions, the modification of the jump rates in
Eq.~\eqref{eq:metropolis-rates} can be written as
\begin{equation}
w_{i\to j}=w_{i\to j}^0\,e^{\beta q\, \vb E\cdot \vb R_{ij}/2}\,,
\label{eq:wij}
\end{equation}
where $\vb R_{ij}=(\vb R_j\!-\!\vb R_i)$ is the nearest-neighbour vector
pointing from the position $\vb R_i$ of site $i$ to the position $\vb
R_j$ of site $j$. As we are using a regular lattice of sites, where
the disorder in the site positions shall be taken into account by the
noise term in Eq.~\eqref{eq:epsi}, it holds $\vb E\cdot \vb R_{ij}=0$
for jumps orthogonal to the field direction (assumed to be parallel to
one lattice axis).  For jumps in and against the field direction $\vb
E\cdot \vb R_{ij}=\pm Ea$, where $a=|\vb R_{ij}|$ is the lattice
constant that represents the mean jump distance of the mobile ions in
the glassy network ($E=|\vb E|$).  Relevant for the conductivity, is
the behaviour in the linear response limit, where Eq.~\eqref{eq:wij}
gives $w_{i\to j}=w_{i\to j}^0(1+\beta q\, \vb E\cdot \vb
R_{ij}/2)$.\footnote{What is truly relevant for the following theory too be valid
  is that the field-perturbed rates satisfy $w_{i\to
    j}/w_{j\to i}=[w_{i\to j}^{(0)}/w_{j\to i}^{(0)}](1+\beta q\vb
  E\cdot\vb R_{ij}$). This can be interpreted as a kind of local
  detailed balance condition in the linear response limit.}

The activation energy for the dc-conductivity can be calculated by
analytical means by applying the theory developed by Ambegaokar,
Halperin and Langer (AHL theory) \cite{Ambegaokar/etal:1971}. In
addition to the field perturbation of the hopping rate in the linear
response limit, the AHL theory takes into account also the change of
the mean occupation numbers of the ion sites from the equilibrium to
the nonequilibrium steady state, where a current is flowing. It maps
the many-body hopping dynamics to that of a disordered conductance
network with links between the ion sites.

The conductance of the link between the sites $i$ and $j$ is given by
\begin{equation}
g_{ij}=\beta q^2w_{i\to j}^0\langle n_i\rangle_{\rm eq}(1-\langle n_j\rangle_{\rm eq})=g_{ji}\,,
\label{eq:gij}
\end{equation}
where the symmetry $g_{ij}=g_{ji}$ follows from the detailed balance
condition of the jump rates $w_{i\to j}^0$ in the absence of the
electric field [see the discussion after Eq.~\eqref{eq:nieq}]. With
the rates $w_{i\to j}^0$ from Eq.~\eqref{eq:metropolis-rates}, these
conductances become
\begin{subequations}
\label{eq:gij-lowT}
\begin{align}
g_{ij}&\sim\beta q^2\nu\exp(-\beta\Delta_{ij})\,,\label{eq:gij-lowT-a}\\
\Delta_{ij}&=u_0+\frac{1}{2}\left(|\varepsilon_i-\varepsilon_j|
+|\varepsilon_i-\varepsilon_{\rms F}|+|\varepsilon_j-\varepsilon_{\rms F}|\right)
\label{eq:gij-lowT-b}
\end{align}
\end{subequations}
in the low-temperature limit $T\to0$. Accordingly, one can view the
$\Delta_{ij}$ as effective barriers for a one-particle hopping in an
energy landscape with equal site energies.

The conductivity activation energy of the conductance network can be
calculated by using percolation theory. It is given by the critical
percolation barrier $\Delta_{\rm c}$. Considering the set of links
$(ij)$ with $\Delta_{ij}$ smaller than a value $\Delta$, the critical
value $ \Delta_{\rm c}$ is the minimal $\Delta$, where neighbouring
(connected) links with $\Delta_{ij}\le\Delta$ form a long-range
percolating path (incipient infinite percolation cluster of links),
\begin{equation}
E_{\rm a}=\Delta_{\rm c}=\min_\Delta\left\{
\Delta\,\Biggl|\, \parbox[c]{15em}{the set of links $(ij)$ with $\Delta_{ij}\le\Delta$\\ forms a percolating path}
\right\}\,.
\label{eq:ea}
\end{equation}
We obtain $\Delta_{\rm c}$ by generating site energy landscapes as
described in Sec.~\ref{sec:site-energies}, then assign the
$\Delta_{ij}$ to the links $(ij)$ between the ion sites according to
Eq.~\eqref{eq:gij-lowT-b}, and eventually determine $\Delta_{\rm c}$
by employing the Hoshen- Kopelmann algorithm
\cite{Hoshen/Kopelman:1976}. The lattices in our analysis consisted of
$N_{\rm tot}=N_L\times N_L\times N_L$ sites with $N_L=100$
corresponding to a linear system size $L=N_La$.

\begin{figure*}[t!]
\centering
\includegraphics[width=0.7\textwidth]{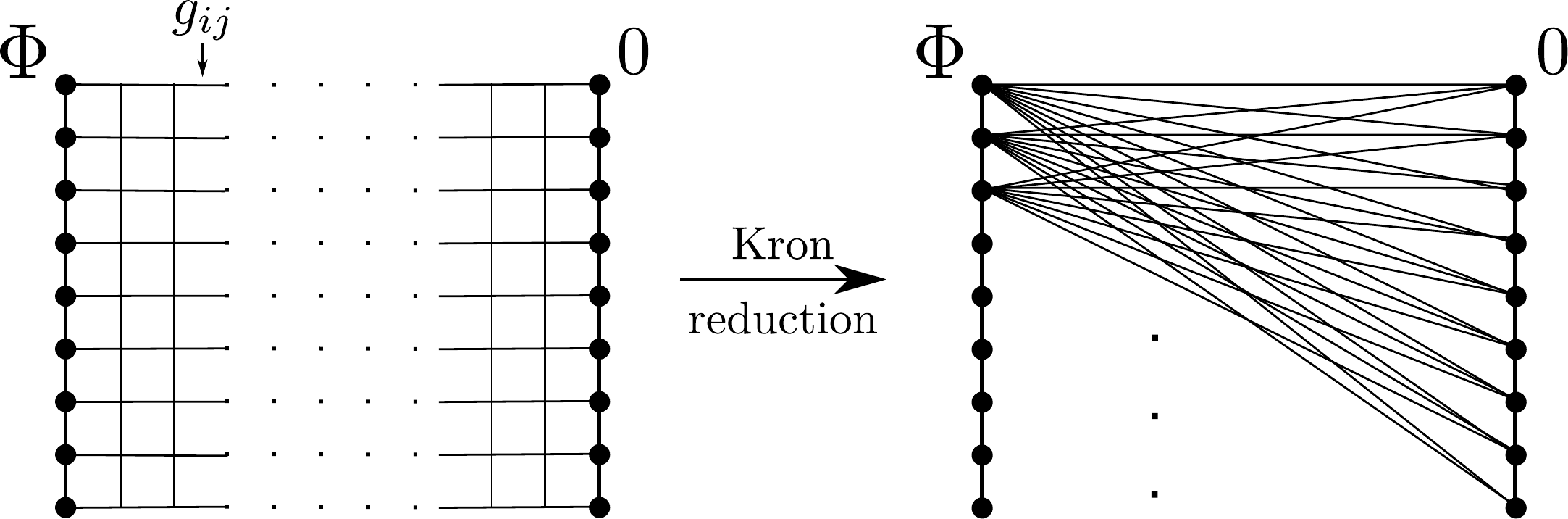}
\caption{Illustration of the Kron reduction for calculating the total
  conductance of a network of nodes $i$ with links having conductances
  $g_{ij}$ (here a regular lattice of ions sites). The passive nodes
  in the interior of the lattice are eliminated after Kron reduction
  and only the active nodes at the boundaries are becoming connected,
  which are at constant potentials $\phi$ and zero.  The effective
  conductances between the active nodes in the reduced network are
  given by the matrix elements of $\bm G$ in Eq.~\eqref{eq:G}.}
\label{fig:kron-reduction}
\end{figure*}

Because of its cubic geometry, the total conductance $G_{\rm tot}$ of the model system
is related to the conductivity by $G_{\rm tot}=\sigma_{\rm dc}L^2/L=L\sigma_{\rm dc}$.
To calculate $G_{\rm tot}$ of the lattice (network) with the link
conductances $g_{ij}$ from Eq.~\eqref{eq:gij}, we 
consider all sites (nodes) belonging to one boundary lattice
plane to be at a constant potential $\phi>0$, and all sites (nodes) at
the opposing lattice plane to be at zero potential, see
Fig.~\ref{fig:kron-reduction}. This corresponds to applying a voltage
$\phi$ to the system.  The total current $I_{\rm tot}$ through the
system is given by $I_{\rm tot}=G_{\rm tot}\phi$, where $G_{\rm tot}$
is calculated by solving Kirchhoff's equations for the current flow in
the network.

Specifically, we introduce the set $\mathcal{B}_+$ of all nodes at the
boundary with potential $\phi$, the set $\mathcal{B}_-$ of all nodes
at the boundary with zero potential, and the set $\mathcal{L}$ of all
other nodes belonging to the interior of the system. For an interior node $i$ 
without connection to the boundary, it holds $\sum_{j} g_{ij}(\phi_j-\phi_i)=0$, 
where the sum over $j$  runs over all nearest neighbours of node $i$ 
(the sum of the currents leaving the node $i$ must be zero due to charge conservation).
For a node $i$ belonging to $\mathcal{B}_+$, Kirchhoffs's law is simply 
$I_i^+=g_{ij}(\phi_j-\phi)$, where $j$ is the (unique) node $j\in\mathcal{L}$ 
linked to the node $i$. Analogous equations hold for nodes belonging to 
$\mathcal{B}_-$ and nodes belonging to $\mathcal{L}$ that are linked to 
boundary nodes. 

These linear equations between the local currents and local potentials can be written in a vector-matrix notation,
\begin{equation}
\begin{pmatrix}
\bm I_+ \\ \bm I_- \\ \bm 0_{\mathcal{L}}
\end{pmatrix}
=
\begin{pmatrix}
\bm G_{++} & \bm 0_{\mathcal{B}} & \bm G_{+\mathcal{L}}\\
\bm 0_{\mathcal{B}} & \bm G_{--} & \bm G_{-\mathcal{L}}\\
\bm G_{\mathcal{L}+} & \bm G_{\mathcal{L}-} & \bm G_{\mathcal{L}\mathcal{L}}\\
\end{pmatrix}
\begin{pmatrix}
\bm \phi \\ \bm 0_- \\ \bm\phi_{\mathcal{L}}
\end{pmatrix}\,,
\label{eq:I-phi}
\end{equation}
where $\bm I_+$ and $\bm I_-$ are the vectors of currents leaving the
nodes of $\mathcal{B}_+$ and $\mathcal{B}_-$ (the vector elements of
$\bm I_-$ being negative because currents are flowing into the nodes
of $\mathcal{B}_-$), and $\bm 0_{\mathcal{L}}$ is the vector of total
currents leaving nodes $i\in\mathcal{L}$ (these currents are all zero,
as discussed above).  The $\bm\phi$ and $\bm 0_-$ are the vectors of
potentials at the nodes of $\mathcal{B}_+$ and $\mathcal{B}_-$ (all
components of $\bm\phi$ are equal to $\phi$ and all components of $\bm
0_-$ are zero), and $\bm\phi_{\mathcal{L}}$ is the vector of
potentials at the nodes $i\in\mathcal{L}$. The block matrices $\bm
G_{\ldots}$ connect the currents and potentials according to
Kirchhoff's law as described above. Their elements are either zero or
equal to link conductances $\pm g_{ij}$, or they are equal to sums of
link conductances (for diagonal elements).  The block matrices $\bm
0_{\mathcal{B}}$ appear, because the nodes of $\mathcal{B_+}$ and
$\mathcal{B}_-$ are not linked.\footnote{As for the dimensions of the
  vectors they are: $N_L^2$ for $\bm I_+$, $\bm I_-$, $\bm\phi$ and
  $\bm 0_-$ and $(N_L^3-2N_L^2)$ for $\bm 0_{\mathcal{L}}$. For the
  matrices the dimensions are $N_L^2\times N_L^2$ for $\bm G_{++}$,
  $\bm G_{--}$, and $\bm 0_\mathcal{B}$, $N_L^2\times (N_L^3-2N_L^2)$
  for $\bm G_{+\mathcal{L}}$ and $\bm G_{-\mathcal{L}}$, and
  $(N_L^3-2N_L^2)\times (N_L^3-2N_L^2)$ for $\bm
  G_{\mathcal{L}\mathcal{L}}$. The matrices $\bm G_{\mathcal{L}+}$ and
  $\bm G_{\mathcal{L}-}$ are the transpose of $\bm G_{+\mathcal{L}}$
  and $\bm G_{-\mathcal{L}}$, respectively.}

From Eq.~\eqref{eq:I-phi}, third row, we obtain $\bm
G_{\mathcal{L}+}\bm\phi+\bm G_{\mathcal{L}\mathcal{L}}
\bm\phi_{\mathcal{L}}=\bm 0_{\mathcal{L}}$.  As $\bm
G_{\mathcal{L}\mathcal{L}}$ is invertible, this yields
$\bm\phi_{\mathcal{L}}=-\bm G_{\mathcal{L}\mathcal{L}}^{-1}\bm
G_{\mathcal{L}+}\bm\phi$.  The first row in Eq.~\eqref{eq:I-phi} then
gives $\bm I_+=\bm G_{++} \bm\phi+\bm
G_{+\mathcal{L}}\bm\phi_{\mathcal{L}}= (\bm G_{++}-\bm
G_{+\mathcal{L}}\bm G_{\mathcal{L}\mathcal{L}}^{-1}\bm
G_{\mathcal{L}+})\bm\phi$.  Accordingly, we can write $\bm I_+=\bm G
\bm\phi$ with
\begin{equation}
\bm G=\bm G_{++}-\bm G_{+\mathcal{L}}\bm G_{\mathcal{L}\mathcal{L}}^{-1}\bm G_{\mathcal{L}+}\,.
\label{eq:G}
\end{equation}
This corresponds to a Kron reduction \cite{Doerfler/Bullo:2013}, where
all ``passive nodes'' (with zero sums of leaving local currents) are
eliminated and only ``active nodes'' remain. In a Kron reduced
network, two active nodes are linked, if a path of linked passive
nodes between them exists in the non-reduced network. In the lattice
considered here, the active nodes are the boundary ones belonging to
$\mathcal{B}_+$ and $\mathcal{B}_-$ and essentially all of them are
becoming connected in the Kron reduced network as indicated in
Fig.~\ref{fig:kron-reduction} (exceptions may be present due to the
presence of the blocked sites).

The total current $I_{\rm tot}$ is equal to the sum of all components
of $\bm I_+$ (and equal to the negative sum of all components of $\bm
I_-$), $I_{\rm tot}=\sum_i (\bm I_+)_i=\sum_{i,j} (\bm
G)_{ij}(\bm\phi)_j=\phi\sum_{i,j} (\bm G)_{ij}$.  The total
conductance of the network is thus given by the sum of all matrix
elements of $\bm G$, and we obtain
\begin{align}
\sigma_{\rm dc}&=\frac{1}{L}G_{\rm tot}=\frac{1}{L}\sum_{i,j} (\bm G)_{ij}\\
&=\frac{1}{L}\sum_{i,j} 
\left(\bm G_{++}-\bm G_{+\mathcal{L}}\bm G_{\mathcal{L}\mathcal{L}}^{-1}\bm G_{\mathcal{L}+}\right)_{ij}\,.
\label{eq:sigmadc}
\end{align}
We calculated this conductivity by applying the so-called method of
iterative Kron reduction \cite{Doerfler/Bullo:2013} to obtain the
matrix elements of $\bm G$.

\section{Activation Energies and Conductivities:\\ Comparison with Experiments}
\label{sec:conductivity-comparison}
We apply the theory described in the previous sections to the three
series of alkali borophophate glasses with compositions
$0.33$Li$_2$O$-0.67$[$x$B$_2$O$_3$-$(1\!-\!x)$P$_2$O$_5$],
$0.35$Na$_2$O$-0.65$[$x$B$_2$O$_3$-$(1\!-\!x)$P$_2$O$_5$], and
$0.4$Na$_2$O$-0.6$[$x$B$_2$O$_3$-$(1\!-\!x)$P$_2$O$_5$] with $0\le
x\le 1$. For these glasses, experimental results for conductivities and
their activation energies were reported in the literature.  The
activation energies have been modelled before \cite{Schuch/etal:2011,
  Bosi/etal:2021}, but in these former studies the barrier $u_0$ (see
Sec.~\ref{sec:hopping-model}) was not taken into account.  Also, we
consider a fixed vacancy fraction $f_0$ of 5\% here for all systems,
while in our former studies results were reported mainly for 10\% of
vacant ion sites. Theoretical predictions for conductivities have not
been compared yet with experimental data. We discuss the results for
activation energies and conductivities separately in the following.

\subsection{Glasses 0.33Li$_\bm{2}$O-0.67[$\bm{x}$B$_\bm{2}$O$_\bm{3}$-$\bm{(1\!-\!x)}$P$_\bm{2}$O$_\bm{5}$]}
\label{subsec:Li33}
\begin{figure}[b!]
\centering
\includegraphics[width=0.7\columnwidth]{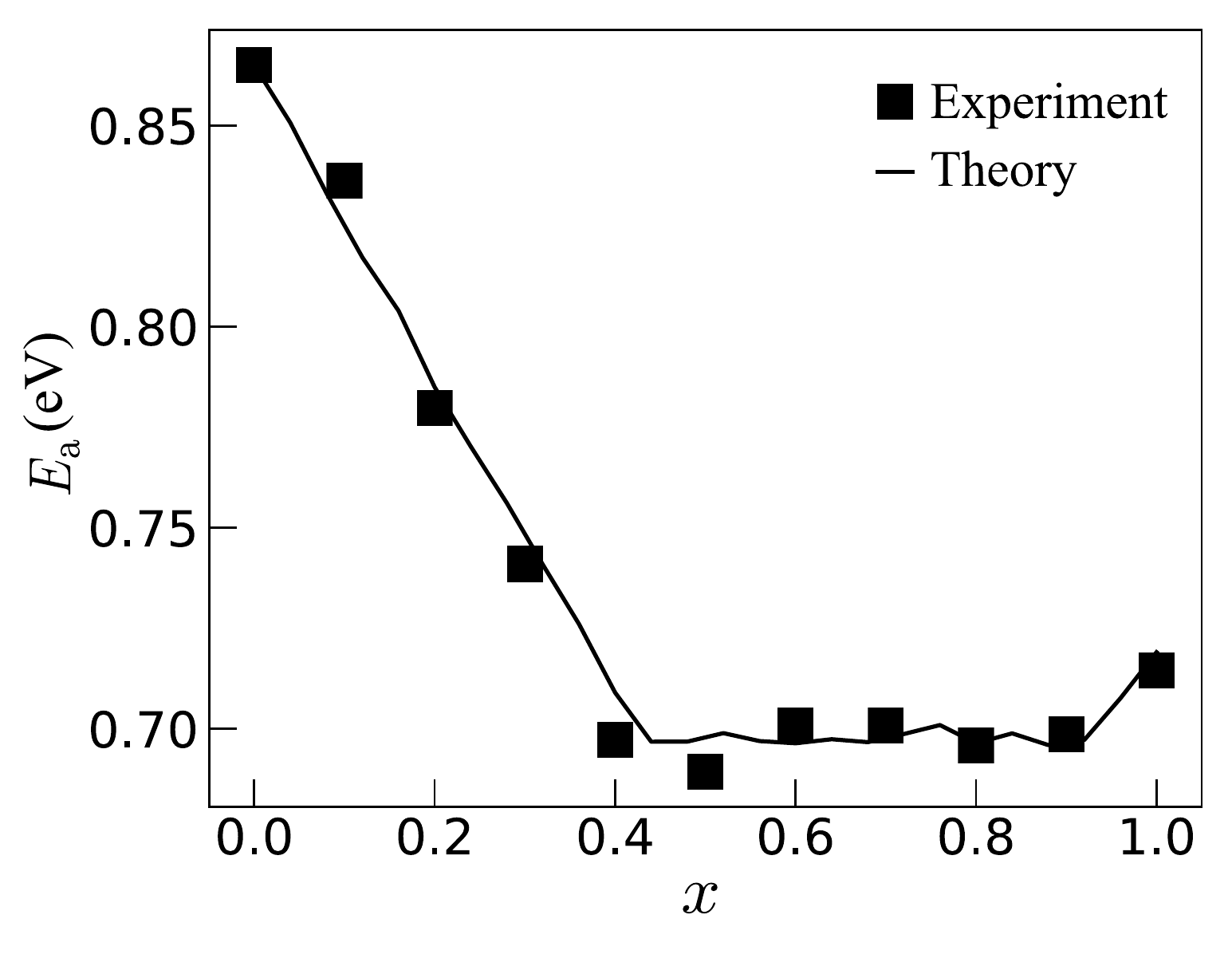}
\vspace{-3ex}
\caption{Conductivity activation energy $E_{\rm a}$ of lithium
  borophosphate glasses with compositions
  0.33Li$_2$O-0.67[$x$B$_2$O$_3$-$(1\!-\!x)$P$_2$O$_5$] as a function
  of the borate to phosphate mixing parameter $x$.  The symbols refers
  to experimental data \cite{Storek/etal:2012} and the line to the
  theoretical modelling for parameters $V_0=0.64\,\si{eV}$ and
  $\sigma_\varepsilon=0.42$. The theoretical activation energies were
  calculated in steps $\Delta x=0.04$ for lattices of linear size
  $L=100\,a$.  Averages were performed over 20 disordered energy
  landscapes (generated as described in
  Sec.~\ref{sec:site-energies}). The relative standard deviation of
  the calculated mean $E_{\rm a}$ values does not exceed 0.4\%.}
\label{fig:EaLi33}
\end{figure}

Figure~\ref{fig:EaLi33} shows the experimental data (symbols) for the
activation energy \cite{Storek/etal:2012} in comparison with the
theoretical modelling (solid line) for the lithium borophosphate
glasses with compositions
0.33Li$_2$O-0.67[$x$B$_2$O$_3$-$(1\!-\!x)$P$_2$O$_5$]. The energy
scale $V_0=0.64\,\si{eV}$ and the noise strength
$\sigma_\varepsilon=0.42$ [see Eq.~\eqref{eq:epsi}] were determined by
requiring the theoretical values at $x=0$ and $x=1$, i.e.\ for the
lithium borate and phosphate glass, to agree with the experimental
data.  As can be seen from the figure, the agreement between theory
and experiment is very good for the mixed glass former glasses with
$x\ne0,1$.

For comparison of conductivity data, we need to specify one further
parameter. According to Eq.~\eqref{eq:sigmadc}, $\sigma_{\rm dc}$ is
proportional to $1/L$ and to the attempt frequency $\nu$, which enters
the local conductances in Eq.~\eqref{eq:gij} via the hopping rates
$w_{i\to j}^0$ in Eq.~\eqref{eq:metropolis-rates}. Since $L=N_La$ and
the number $N_L^3$ of sites known in our modelling, the relevant
parameter is $\nu/a$. The mean jump distance $a$ can be estimated to
be about $5\si{\AA}$ in the alkali borophosphate glasses, see, for
example, Fig.~11 in Ref.~\cite{Christensen/etal:2013}. We fix this
value for $a$ (independent of $x$), although the mean jump distance
can be expected to vary slightly with $x$. Indeed a weak decrease of
$a$ with $x$ has been reported in
Ref.~\cite{Christensen/etal:2013}. We note that also the other
parameters $V_0$ and $\sigma_\varepsilon$ should vary with $x$, but our
successful modelling of the activation energies with fixed values of
$V_0$ and $\sigma_\varepsilon$ suggests that the variation is weak also.

\begin{figure}[b!]
\centering
\includegraphics[width=0.7\columnwidth]{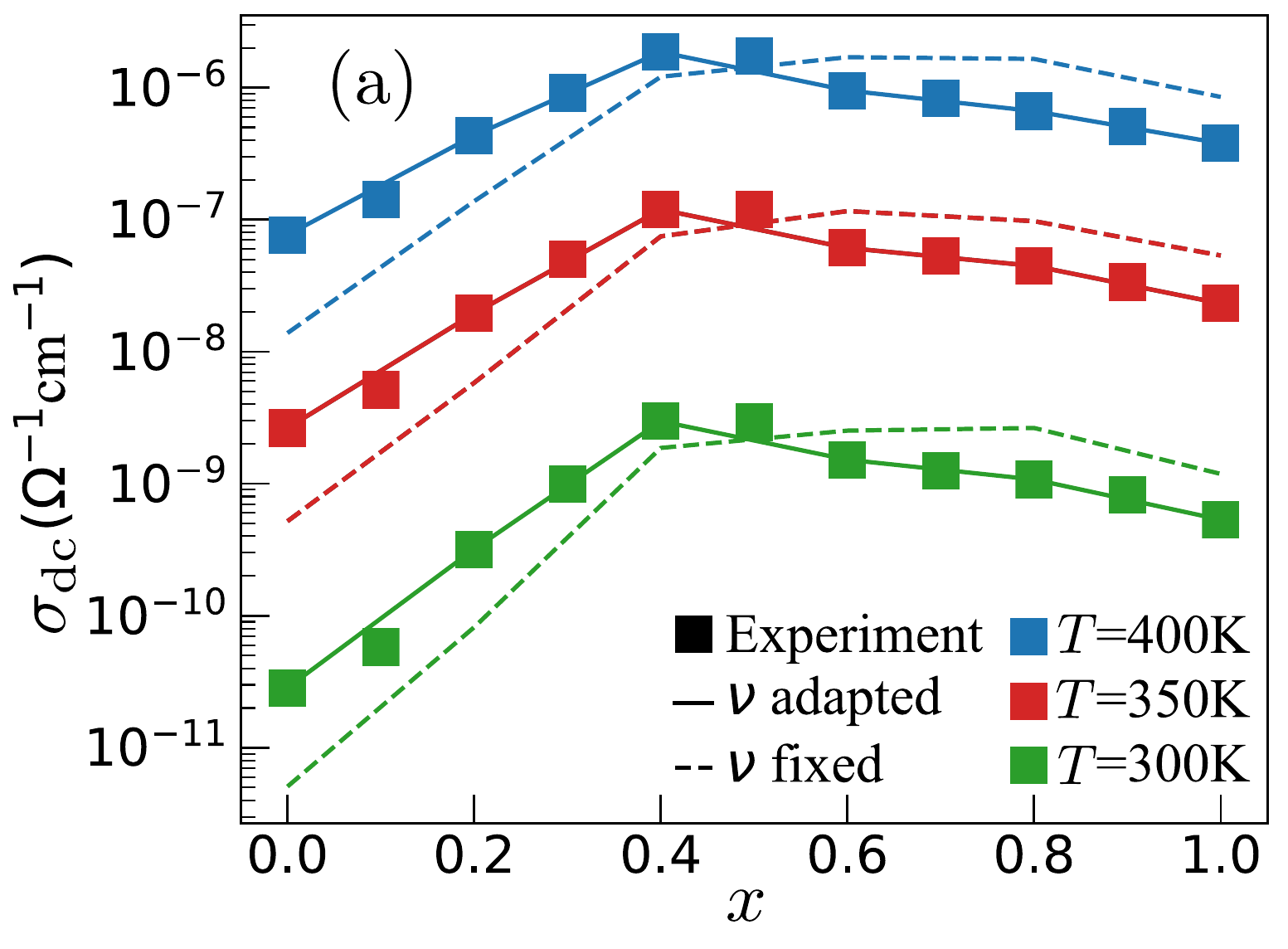}
\hspace*{2em}\includegraphics[width=0.66\columnwidth]{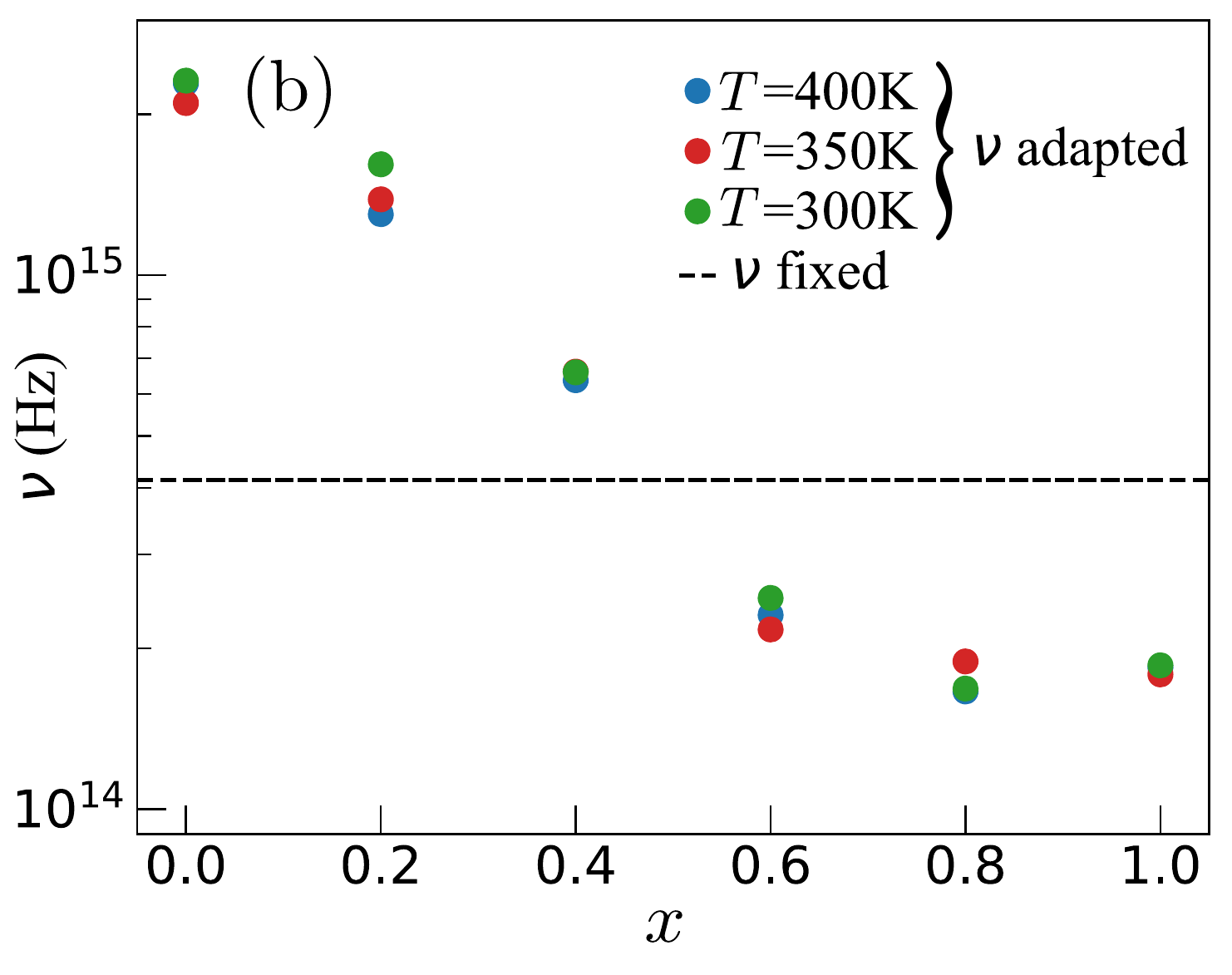}
\vspace{-3ex}
\caption{(a) Conductivity $\sigma_{\rm dc}$ of lithium borophosphate
  glasses with compositions
  0.33Li$_2$O-0.67[$x$B$_2$O$_3$-$(1\!-\!x)$P$_2$O$_5$] as a function
  of $x$.  The symbols refer to experimental data
  \cite{Storek/etal:2012} and the lines to theoretical calculations
  for the same parameters $V_0$ and $\sigma_\varepsilon$ as in
  Fig.~\ref{fig:EaLi33}.  The dashed and solid lines refer to the
  theoretical modellings with a fixed and with adapted attempt
  frequencies $\nu$, respectively. (b) Attempt frequencies $\nu$ used
  for fitting of the conductivity data.  The horizontal dashed line
  marks the fixed, $x$-independent $\nu$ value.  The calculations were
  carried out in steps of $\Delta x=0.2$ for a linear system size
  $L=35\,a$. Averages were performed over 30 energy landscapes. The
  relative standard deviation of the calculated mean $\sigma_{\rm dc}$
  values does not exceed 15\%.}
\label{fig:CondLi33}
\end{figure}

More difficult is to make reasonable assumptions for $\nu$ apart
from estimating it to be of the order of optical phonon frequencies.  The
attempt frequency in particular contains entropic contributions,
including a ``local migration entropy'' not considered in our
coarse-grained approach based on a hopping model. This local migration
entropy quantifies the mean number of relevant pathways (or typical
configuration space) accessible for a mobile ion in a rare transition from one ion site to a neighboring one, 
which, on a coarse-grained time scale, is described as an instantaneous hopping process.

In view of this complexity, we use two procedures to specify $\nu$. In
the first one, we consider also $\nu$ to have a fixed value in order
to keep the number of model parameters as small as possible. An
interesting question then is, if we can successfully recover the
overall change of conductivities with $x$ as found in the
experiments. In the second procedure, we adjust $\nu$ to fit the
experimental conductivity data for each value of $x$. This can be
considered as an ``overfitting'', because irrespective of our
construction of the underlying energy landscape described in
Sec.~\ref{sec:site-energies}, it would be always possible to match the
measured data. However, the adapted attempt frequencies should have
reasonable values and their variation can provide interesting insights
into entropy variations.

Figure~\ref{fig:CondLi33}(a) shows the experimental data for the
conductivity (symbols) for three temperatures in comparison with the
theoretical modelling for fixed $\nu$ (dashed lines) and adapted $\nu$
(solid lines). The corresponding attempt frequencies are displayed in
Fig.~\ref{fig:CondLi33}(b), where the dashed horizontal line marks the
fixed, $x$-independent $\nu$ value. From the comparison of the
theoretical with the experimental results in Fig.~\ref{fig:CondLi33}(a), we
can conclude that the overall variation of $\sigma_{\rm dc}(x)$ is
captured by the modelling with a fixed attempt frequency $\nu$, but a
quantitative agreement is not obtained.  The value $\nu=4\times
10^{14}\,\si{Hz}$, found by least squared error fitting of the measured data,
has a reasonable order of magnitude.

When adjusting $\nu$ to match the measured data, we find a decrease of
the adapted $\nu$ values from values of about $10^{15}\,\si{Hz}$ to
values of about $10^{14}\,\si{Hz}$ when $x$ is increased from zero to
one, see Fig.~\ref{fig:CondLi33}(b). All these values are of
reasonable order of magnitude. Interestingly, a similar variation of
pre-exponential factors from $10^{16}\,\si{Hz}$ down to
$10^{14}\,\si{Hz}$ has been found by fitting NMR spectra
\cite{Storek/etal:2012}.

\subsection{Glasses 0.35Na$_\bm{2}$O-0.65[$\bm{x}$B$_\bm{2}$O$_\bm{3}$-$\bm{(1\!-\!x)}$P$_\bm{2}$O$_\bm{5}$]}
\label{subsec:Na35}

\begin{figure}[b!]
\centering \includegraphics[width=0.8\columnwidth]{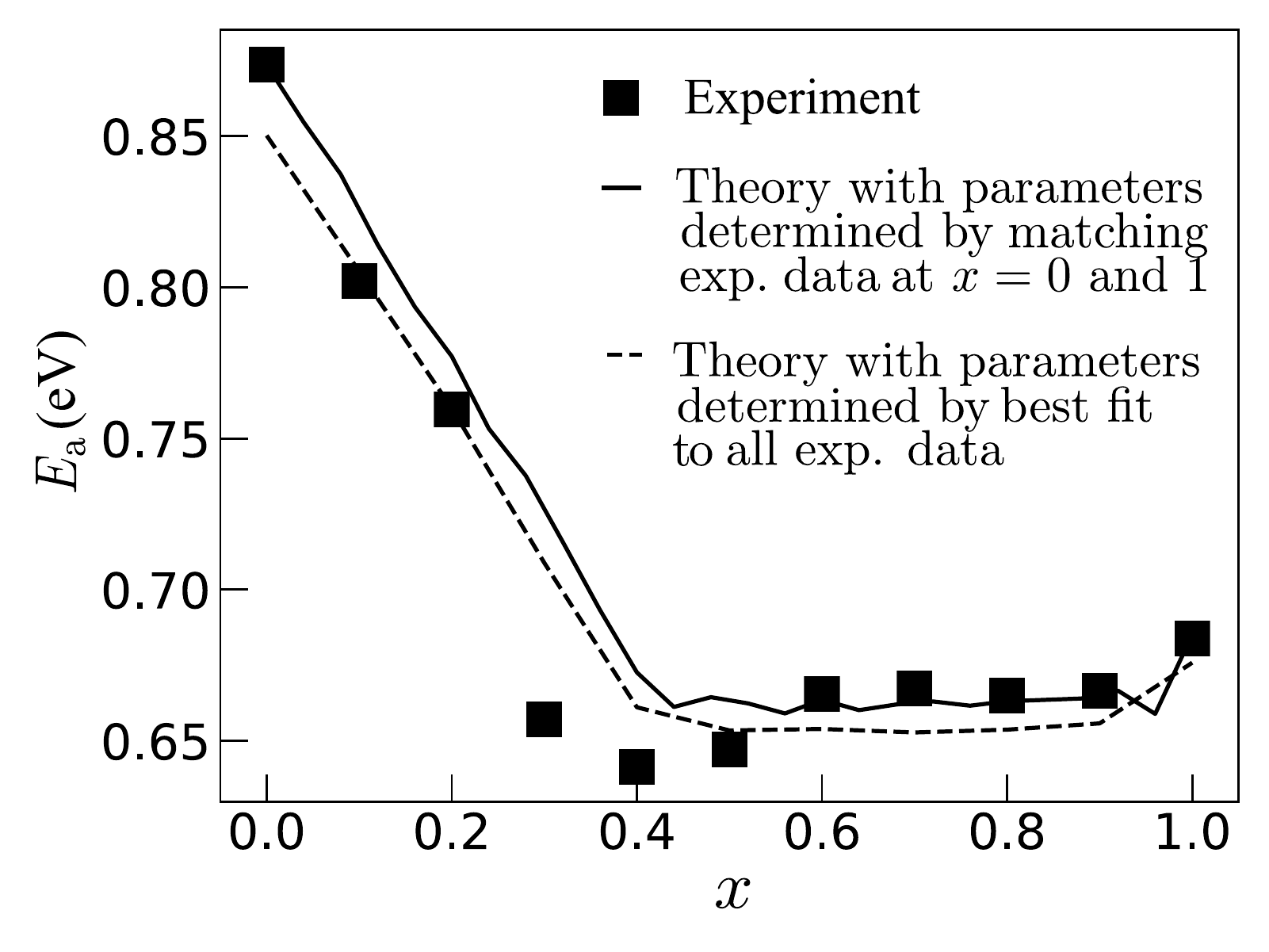}
\vspace{-3ex}
\caption{Conductivity activation energy $E_{\rm a}$ of sodium
  borophosphate glasses with compositions
  0.35Na$_2$O-0.65[$x$B$_2$O$_3$-$(1\!-\!x)$P$_2$O$_5$] as a function
  of the borate to phosphate mixing parameter $x$.  The symbols refer
  to experimental data \cite{Christensen/etal:2013} and the lines to the
  theoretical modelling for $V_0=0.76\,\si{eV}$ and
  $\sigma_\varepsilon=0.30$ (solid line), and for $V_0=0.73\,\si{eV}$ and
  $\sigma_\varepsilon=0.32$ (dashed line). The step size $\Delta x$,
  system size and disorder averaging used in the calculations are the
  same as in Fig.~\ref{fig:EaLi33}. The relative standard deviation of
  the calculated mean $E_{\rm a}$ values does not exceed 0.5\% (for
  both modellings with the different parameter values).}
\label{fig:EaNa35}
\end{figure}

\begin{figure}[t!]
\centering
\includegraphics[width=\columnwidth]{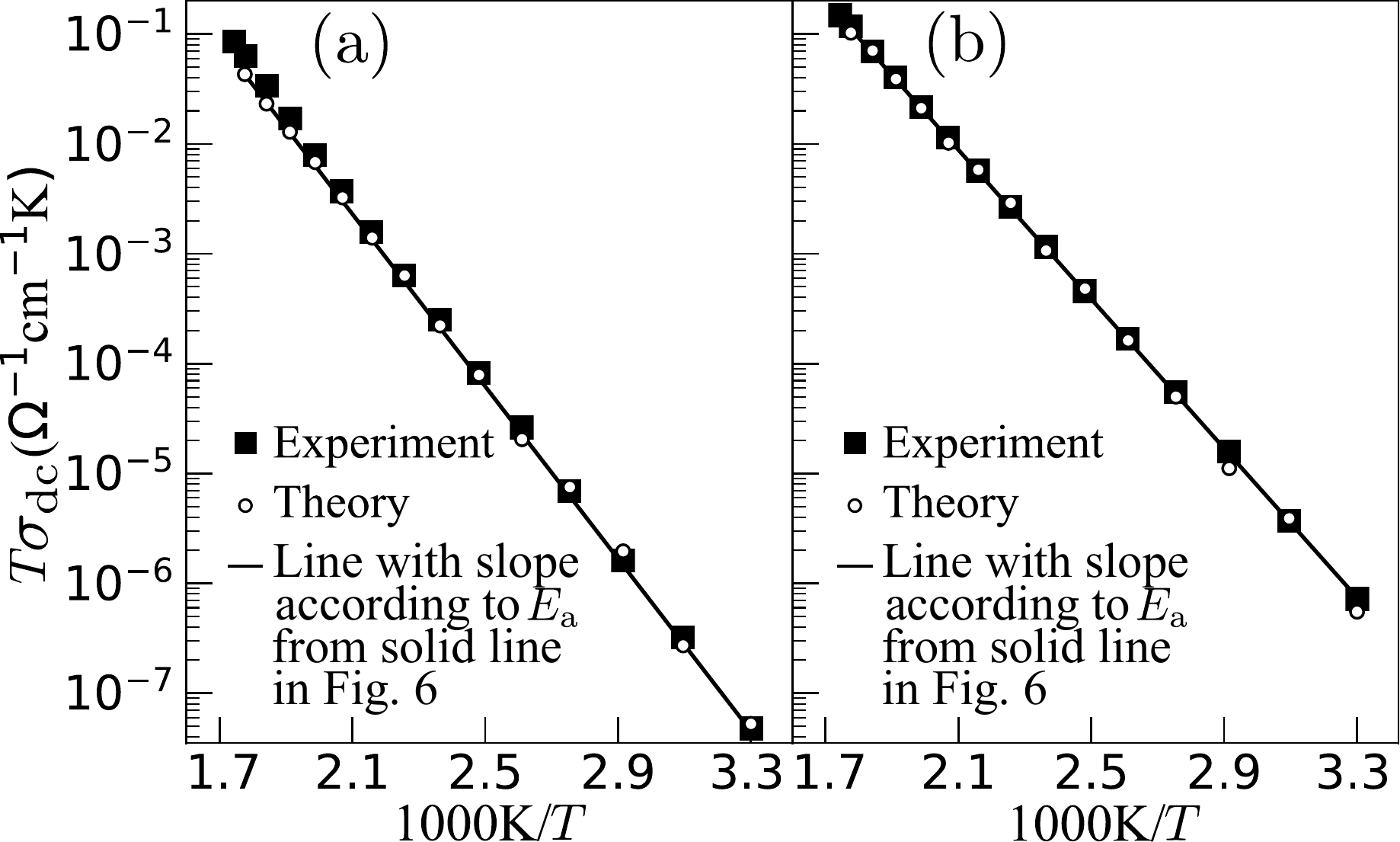}
\vspace{-4ex}
\caption{Arrhenius plots of $T\sigma_{\rm dc}$ in sodium borophosphate
  glasses of compositions (a)
  0.35Na$_2$O-0.65[$0.2$B$_2$O$_3$-$0.8$P$_2$O$_5$] ($x=0.2$) and (b)
  0.35Na$_2$O-0.65[$0.4$B$_2$O$_3$-$0.6$P$_2$O$_5$] ($x=0.4$). The
  filled symbols refer to experimental data
  \cite{Christensen/etal:2013} and the open symbols to a modelling for
  parameters $V_0=0.76\,\si{eV}$ and $\sigma_\varepsilon=0.3$,
  i.e.\ equal to the ones yielding the solid line in
  Fig.~\ref{fig:EaNa35}.  The attempt frequencies in the calculations
  are $\nu=5\times10^{14}\,\si{Hz}$ in (a) and
  $\nu=1.5\times10^{14}\,\si{Hz}$ in (b).  The solid lines in (a) and
  (b) have slopes corresponding to the calculated activation energies
  in Fig.~\ref{fig:EaNa35}.  Values of $T\sigma_{\rm dc}$ were
  calculated in the interval $303\,\si{K}$-$563\,\si{K}$ in
  temperature steps of $20\,\si{K}$, and the system size and disorder
  averaging used in the calculations are the same as in
  Fig.~\ref{fig:CondLi33}. The relative standard deviation of the
  calculated mean $\sigma_{\rm dc}$ values does not exceed 23\% in (a)
  and (b).}
\label{fig:ArrhNa35}
\end{figure}

Figure~\ref{fig:EaNa35} shows the comparison of the experimental
\cite{Christensen/etal:2013} (symbols) and modelled conductivity
activation energies (solid and dashed lines) for the sodium
borophosphate glasses with compositions
0.35Na$_2$O-0.65[$x$B$_2$O$_3$-$(1\!-\!x)$P$_2$O$_5$]. When
determining the parameters $V_0$ and $\sigma_\varepsilon$ as for the
lithium borophosphate glasses in the previous Sec.~\ref{subsec:Li33},
i.e.\ by requiring the theoretical values to match the experimental
ones for $x=0$ and $x=1$, we obtain $V_0=0.76\,\si{eV}$ and
$\sigma_\varepsilon=0.30$. The predicted behaviour for $0<x<1$ is given
by the solid line in Fig.~\ref{fig:EaNa35}. The agreement with the
experimental data is satisfactory but less good for $x\lesssim0.5$
than in Fig.~\ref{fig:EaLi33}. As we do not know whether the
experimental data for $x=0$ and $x=1$ have less uncertainty than those
for $x\ne0,1$, we determined the parameters $V_0$ and
$\sigma_\varepsilon$ alternatively also by a least squared error fitting
of the experimental data. This gives $V_0=0.73\,\si{eV}$ and
$\sigma_\varepsilon=0.32$. The corresponding results are shown as dashed
line in Fig.~\ref{fig:EaNa35}. They give a slightly better agreement
for $x\lesssim0.5$, but underestimate the experimental $E_{\rm a}$
values for $x\ge0.6$.

Moreover, we find that Arrhenius plots of measured conductivity data
for $x=0.2$ and $x=0.4$ (filled symbols) can be well described by the
hopping model (open symbols) with the parameters $V_0=0.76\,\si{eV}$
and $\sigma_\varepsilon=0.30$, see Fig.~\ref{fig:ArrhNa35}. The solid
lines in Fig.~\ref{fig:ArrhNa35}(a) and (b) have slopes that agree
with the modelled conductivity activation energies given by the solid
line in Fig.~\ref{fig:EaNa35}. This demonstrates that the
corresponding $E_a$ values can account for the experimental findings
also if $x$ is smaller than 0.5. In the further modelling of
dc-conductivities, we use the parameter values $V_0=0.76\,\si{eV}$ and
$\sigma_\varepsilon=0.30$.

\begin{figure}[t!]
\centering
\includegraphics[width=0.7\columnwidth]{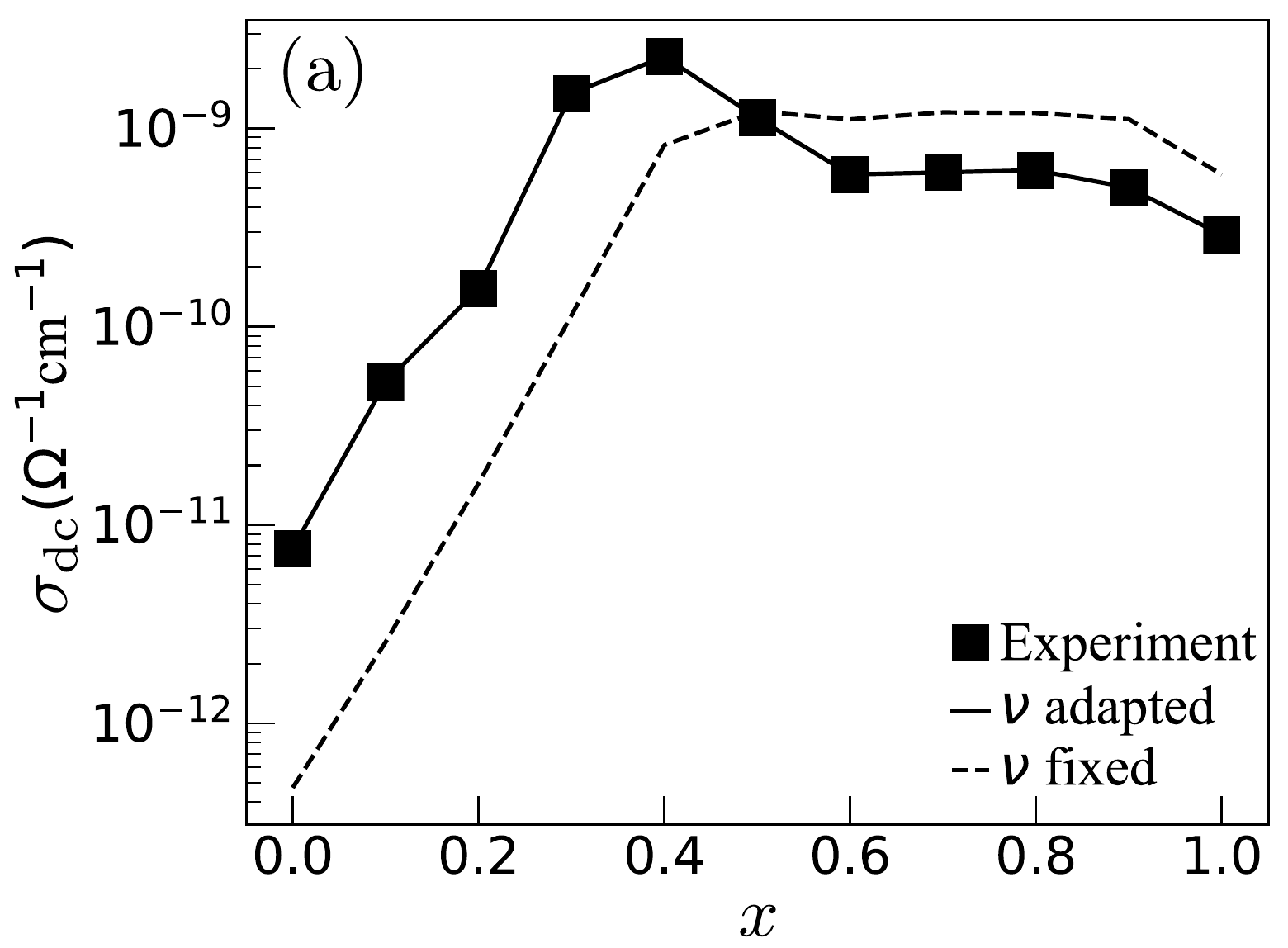}
\hspace*{1em}\includegraphics[width=0.67\columnwidth]{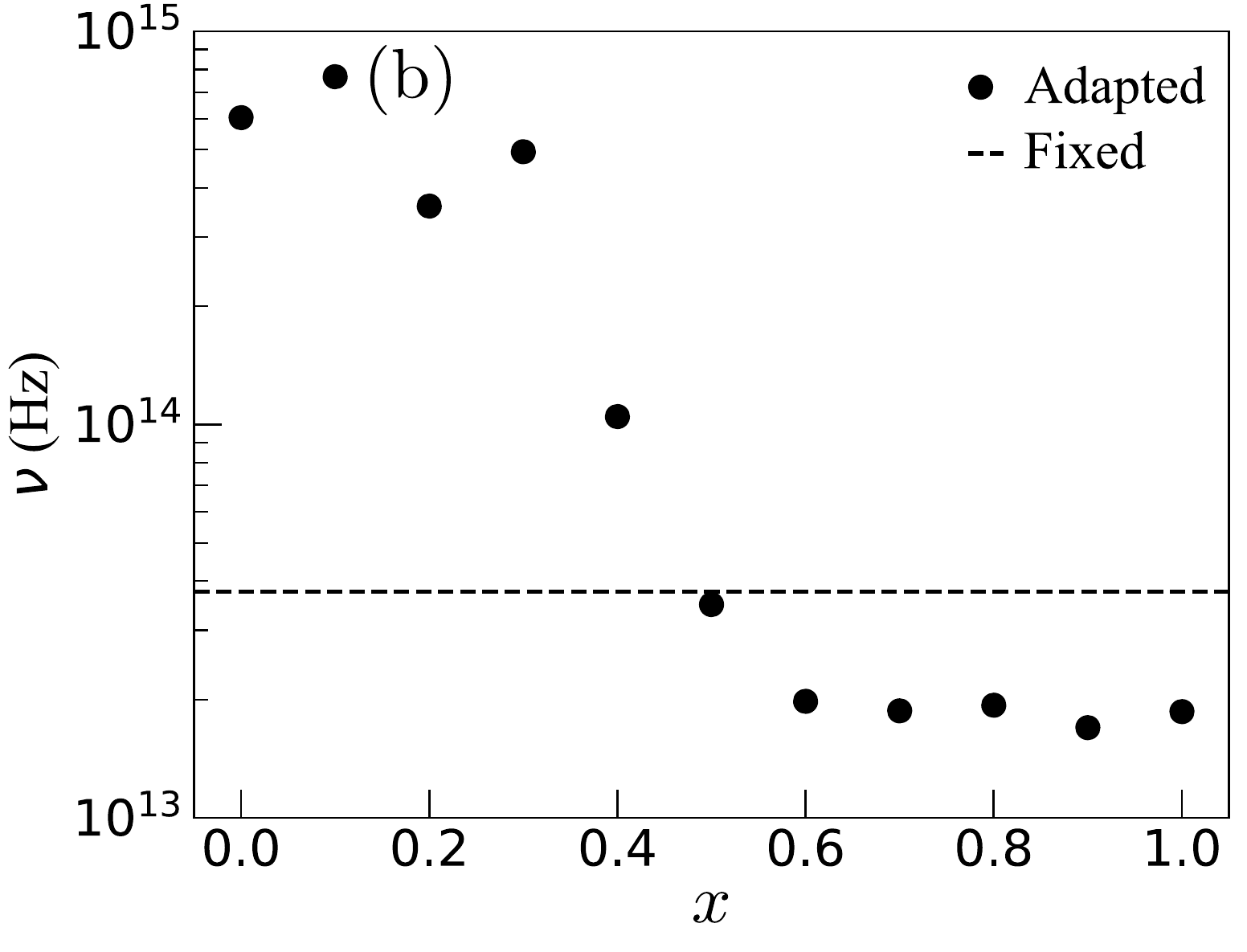}
\vspace{-3ex}
\caption{(a) Conductivity $\sigma_{\rm dc}$ of sodium borophosphate
  glasses with compositions
  0.35Na$_2$O-0.65[$x$B$_2$O$_3$-$(1\!-\!x)$P$_2$O$_5$].  The symbols
  refers to experimental data \cite{Christensen/etal:2013} and the
  lines to theoretical calculations for the parameters
  $V_0=0.76\,\si{eV}$ and $\sigma_\varepsilon=0.30$, i.e.\ corresponding
  to the ones used for the solid line in Fig.~\ref{fig:EaNa35}.  The
  dashed and solid lines refer to the theoretical modelling with a
  fixed and with adapted attempt frequencies $\nu$, respectively. (b)
  Attempt frequencies $\nu$ used for fitting of the conductivity data.
  The horizontal dashed line marks the fixed, $x$-independent $\nu$
  value.  The calculation were performed in steps of $\Delta x=0.1$
  for a system size and a disorder averaging as in
  Fig,~\ref{fig:CondLi33}.  The relative standard deviation of the
  calculated mean $\sigma_{\rm dc}$ values does not exceed 16\%.}
\label{fig:CondNa35}
\end{figure}

Figure~\ref{fig:CondNa35}(a) shows the experimental data for the
conductivity (symbols) for three temperatures in comparison with the
theoretical modelling for fixed $\nu$ (dashed lines) and adapted $\nu$
(solid lines). The corresponding attempt frequencies are displayed in
Figure~\ref{fig:CondNa35}(b). The agreement between experiment and
theory is of similar quality as for the lithium borophosphate glasses
in Fig.~\ref{fig:CondLi33} with attempt frequencies having a
reasonable order of magnitude.

\subsection{Glasses 0.4Na$_\bm{2}$O-0.6[$\bm{x}$B$_\bm{2}$O$_\bm{3}$-$\bm{(1\!-\!x)}$P$_\bm{2}$O$_\bm{5}$]}
\label{subsec:Na40}

\begin{figure}[t!]
\centering
\includegraphics[width=0.8\columnwidth]{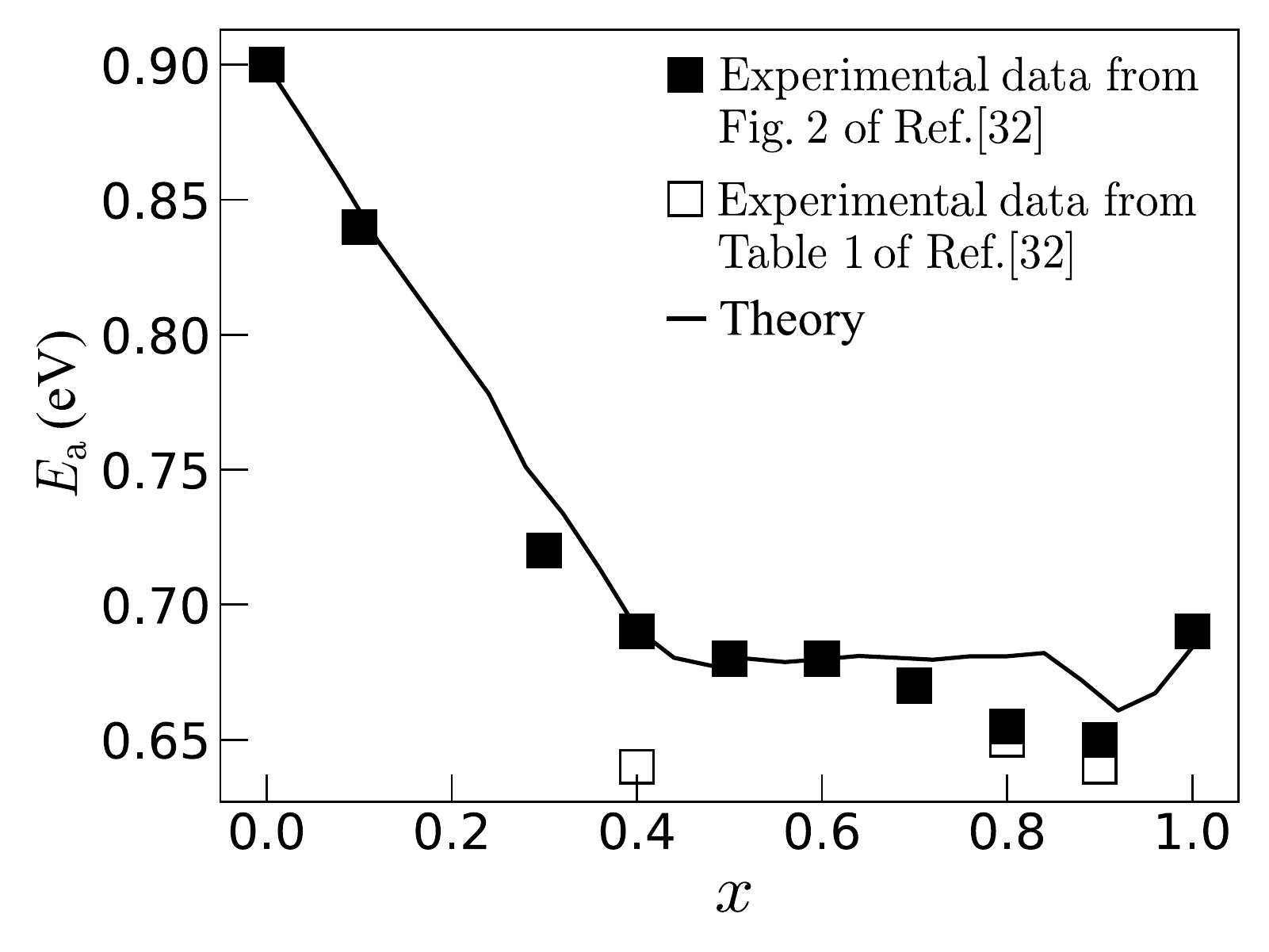}
\vspace{-3ex}
\caption{Conductivity activation energy $E_{\rm a}$ of sodium
  borophosphate glasses with compositions
  0.4Na$_2$O-0.6[$x$B$_2$O$_3$-$(1\!-\!x)$P$_2$O$_5$].  The filled and
  open symbols refer to experimental data taken from table~1 and from
  Fig.~2b) of Ref.~\cite{Zielniok/etal:2007}.  The line refers to the
  theoretical modelling with $V_0=0.93\,\si{eV}$ and
  $\sigma_\varepsilon=0.25$.  The step size $\Delta x$, system size and
  disorder averaging used in the calculations are the same as in
  Figs.~\ref{fig:EaLi33} and \ref{fig:EaNa35}.  The relative standard
  deviation of the calculated mean $E_{\rm a}$ values does not exceed
  0.5\%.\vspace*{0.5ex}
}
\label{fig:EaNa40}
\end{figure}

\begin{figure}[h!]
\centering
\includegraphics[width=0.72\columnwidth]{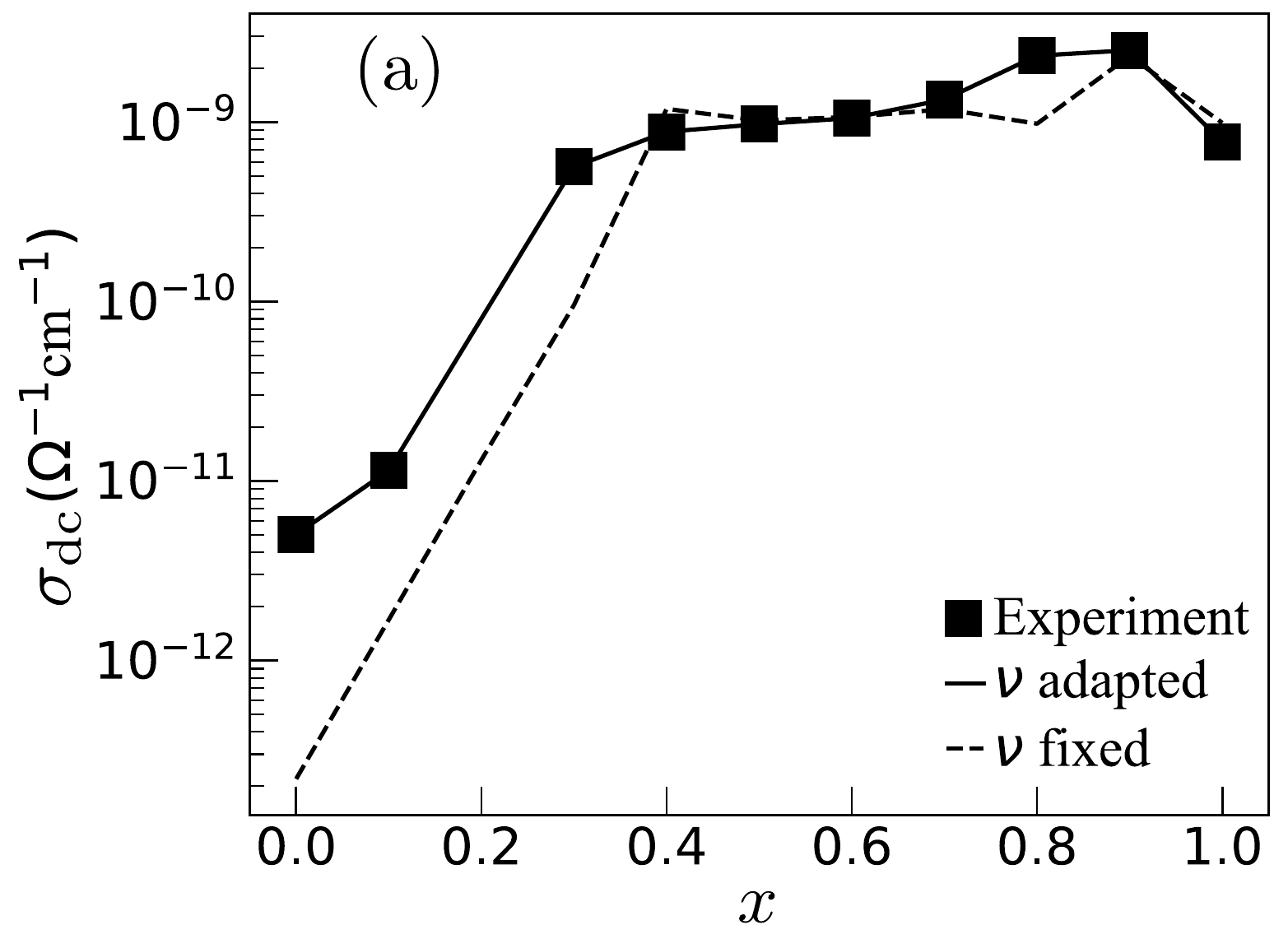}
\hspace*{2em}\includegraphics[width=0.65\columnwidth]{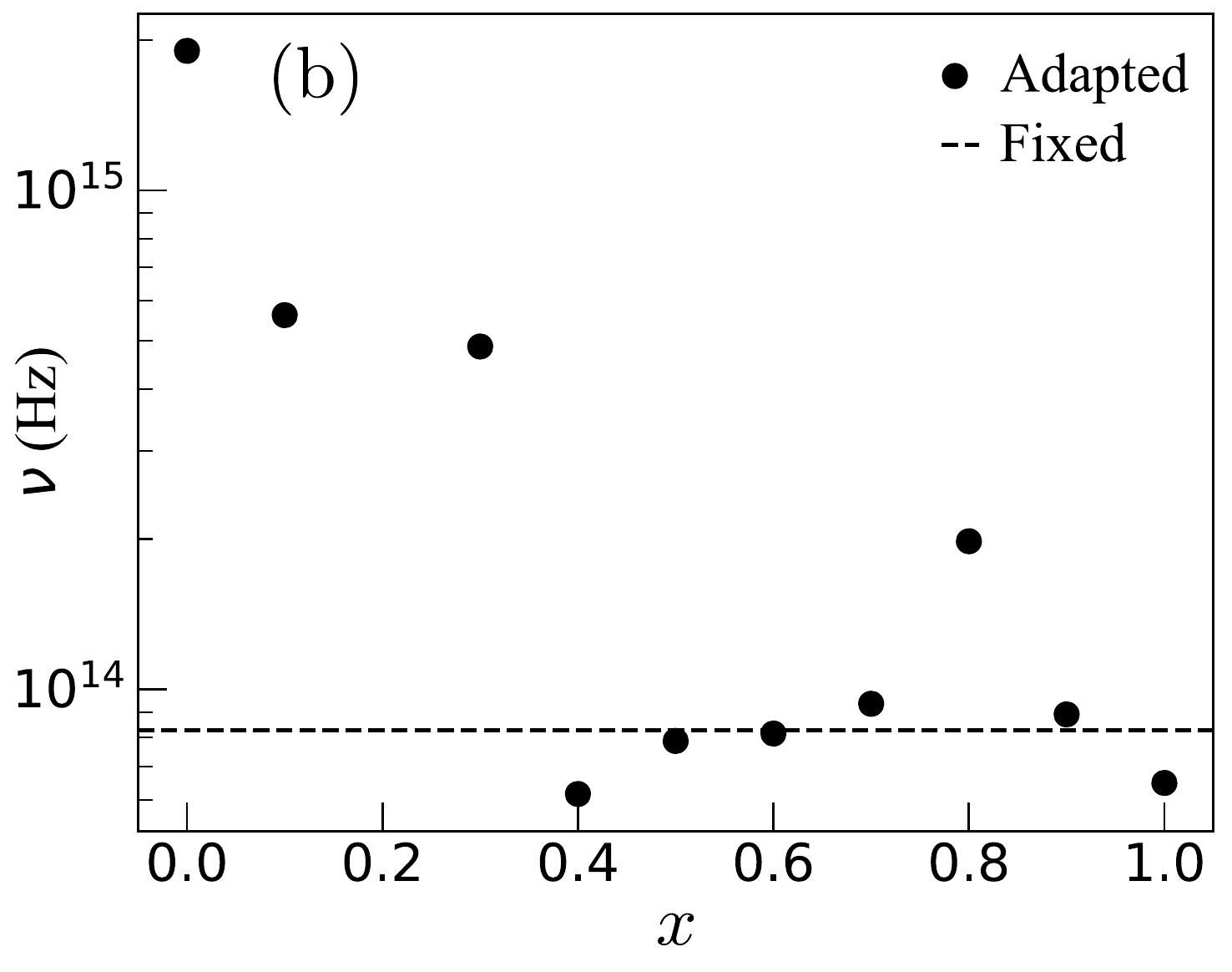}
\vspace{-3ex}
\caption{(a) Conductivity $\sigma_{\rm dc}$ of sodium borophosphate
  glasses with compositions
  0.4Na$_2$O-0.6[$x$B$_2$O$_3$-$(1\!-\!x)$P$_2$O$_5$].  The symbols
  refer to experimental data \cite{Zielniok/etal:2007} and the lines
  to theoretical calculations for the same parameters $V_0$ and
  $\sigma_\varepsilon$ as in Fig.~\ref{fig:EaNa40}.  The dashed and solid
  lines are for the theoretical modelling with a fixed and with
  adapted attempt frequencies $\nu$, respectively. (b) Attempt
  frequencies $\nu$ used for fitting of the conductivity data.  The
  horizontal dashed line marks the fixed, $x$-independent $\nu$ value.
  Step sizes $\Delta x$, linear system size and disorder averaging are
  the same as in Fig.~\ref{fig:CondNa35}.  The relative standard
  deviation of the calculated mean $\sigma_{\rm dc}$ values does not
  exceed 22\%.}
\label{fig:CondNa40}
\end{figure}

Figure~\ref{fig:EaNa40} shows the comparison of the experimental
\cite{Zielniok/etal:2007} (filled and open symbols) with the
calculated conductivity activation energies (solid lines) for sodium
borophosphate glasses with compositions
0.4Na$_2$O-0.6[$x$B$_2$O$_3$-$(1\!-\!x)$P$_2$O$_5$]. The parameters
obtained from matching the measured data at $x=0$ and $x=1$ are
$V_0=0.93\,\si{eV}$ and $\sigma_\varepsilon=0.25$.  As can be seen from
the figure, the theoretical results again agree well with the
experimental findings.

The calculated conductivities $\sigma_{\rm dc}$ in
Fig.~\ref{fig:CondNa40} compare similarly with experimental data as
the other glass series shown in Figs.~\ref{fig:CondLi33} and
\ref{fig:CondNa35}, again with reasonable values of adjusted attempt
frequencies in the range $10^{14}-10^{15}\,\si{Hz}$.

\section{Conclusions}
\label{sec:conclusions}
Ionic conductivities $\sigma_{\rm dc}$ and their activation energies
$E_{\rm a}$ have been successfully modelled for various alkali
borophosphate glasses with compositions
$yM_2$O-$(1\!-\!y)$[$x$B$_2$O$_3$-$(1\!-\!x)$P$_2$O$_5$] ($M$: alkali
ion, $y$: molar fraction of alkali oxide, $x$: borate to phosphate
mixing parameter). This was done based on a construction of site
energy landscapes and a hopping model for the alkali ion motion
developed in earlier studies \cite{Schuch/etal:2011,
  Bosi/etal:2021}. In this model, it was assumed that the Coulomb
interaction can be treated in a mean-field like manner and that it can
be effectively included into a noise contribution to the site
energies, or, more generally speaking, to the jump rates. Whether this
is a valid concept needs to be explored, in particular, whether a Coulomb
gap must be considered in the one-particle density of states, as it
has been studied in the related problem of electron hopping transport in
amorphous semiconductors \cite{Pollak:1970, Efros/Shklovskii:1975,
  Moebius/etal:1992, Mueller/Pankov:2007}.

The energy landscape construction relies on the idea that the
countercharges associated with different types of NFUs are important
and that changes in the concentrations of NFU types dominate the
variation of the energy landscape when the mixing parameter $x$ is
altered. Concentrations of the NFU units can be measured by MAS-NMR
and Raman spectroscopy and they were successfully described
theoretically in the earlier studies by a statistical mechanical
approach \cite{Schuch/etal:2011, Bosi/etal:2021}.  Experimental
activation energies for compositions with different mixing ratio $x$
could be described quantitatively by introducing one energy scale
parameter $V_0$ and a parameter $\sigma_\varepsilon$ characterising
additional noise in the energy landscape.  These two parameters were
determined by fitting of the experimental results for $x=0$ (alkali
borate glass) and $x=1$ (alkali phosphate glass).  The $E_{\rm a}$
values calculated for $0<x<1$ are theoretical predictions then.

A modelling of activation energies has been conducted in the earlier
studies also, first based on KMC simulations \cite{Schuch/etal:2011}
and later by the analytical method used here \cite{Bosi/etal:2021}. In
these former calculations, however, local ion jumps could have
vanishing energetic barriers, which is an unphysical feature. In the
present study we showed that this problem can be resolved when
introducing a minimal barrier $u_0>0$ for local jumps. Let us mention
that a good agreement of the theoretical results with the experimental
observations is not achieved if $u_0$ becomes too large. We estimated
$u_0$ by considering it to have values comparable to the thermal
energy at the glass transition temperature $T_{\rm g}$. A fixed value $u_0$
independent of the borate to phosphate mixing ratio was used for
simplicity. One can refine this approach and adjust $u_0$ to $T_{\rm
  g}$. The change of $T_{\rm g}$ with $x$ in general correlates well
with the number of bridging oxygens, and this number can be calculated
from the known NFU concentrations. It will be interesting to see in
the future, how a corresponding refinement modifies the theoretical
results. In a further refinement one may consider a distribution of
the (lower) barriers for local jumps in addition to the distribution
of ion site energies.

Calculations of conductivities for the hopping model were carried out
here for the first time based on an analytical theoretical method. The
overall variation of measured conductivities was reproduced when
taking an $x$-independent attempt frequency $\nu$ in the jump rates of the
hopping model. When allowing for a variation of the attempt
frequencies with $x$, a quantitative agreement with experimental
results was obtained for reasonable values of $\nu$ varying in
the range $10^{13}-10^{14}\,\si{Hz}$. In the corresponding
calculations we considered a mean jump distance of
$5\,$\AA\ independent of $x$. Generally, the mean jump distance, the
energetic scale parameter $V_0$ and the noise strength
$\sigma_\varepsilon$ can be expected to vary with $x$.  However, in order
to see whether the theory has predictive power, we kept the number of
parameters as small as possible. It is our hope to determine or
estimate the parameters $V_0$ and $\sigma_\varepsilon$ without a fitting
in the future by resorting on additional structural information as,
e.g., gained from RMC modelling and molecular dynamics simulations.

For extending and refining the modelling, it will be furthermore
important to check theoretical predictions for other dynamical probes
commonly studied in experiments, such as ac-conductivities and spin
lattice relaxation rates \cite{Heitjans/etal:2005,
  VinodChandran/Heitjans:2016}. This can be done, in principle, by
extensive kinetic Monte Carlo simulations
\cite{Rinn/etal:1998-p}. Whether corresponding calculations can be
carried out in a reliable manner also by fast numerical solutions of
equations, as done here for dc-conductivities and their activations
energies, is an open problem. A recently developed experimental
technique is the charge attachment induced ion transport (CAIT)
\cite{Schaefer/Weitzel:2018, Schaefer/etal:2019, Weitzel:2021}, where
collective diffusion coefficients of mobile ions are extracted by
a modelling of diffusion profiles. A particular merit of the CAIT method is that mobile
ions of different types can be directly injected in the glassy phase.
This makes it possible that the observed concentration-dependent
diffusion coefficients are governed by other intervals of the site
energy distribution than those governing dc-conductivities
obtained from impedance spectroscopy. Comparison of CAIT results with
theoretical calculations will thus provide an important additional
means for testing the energy landscape construction.

\vspace{2ex}\noindent \textbf{Acknowledgement}\\ 
This work has been funded by the Deutsche Forschungsgemeinschaft (DFG,
Project No.~428906592).  We sincerely thank the members of the DFG
Research Unit FOR 5065 for fruitful discussions.

\end{document}